# Double-domed temperature-pressure phase diagram found for CePd$_3$S$_4$

S. Huyan[1,2], T. J. Slade[1,2], H. Wang[3], R. Flint[1,2], R. A. Ribeiro[1,2], W. Xie[3], S. L. Bud'ko[1,2], P. C. Canfield[1,2]

[1]*Ames National Laboratory, US DOE, Iowa State University, Ames, Iowa 50011, USA*
[2]*Department of Physics and Astronomy, Iowa State University, Ames, Iowa 50011, USA*
[3] *Department of Chemistry, Michigan State University, East Lansing, MI 48824, USA*

e-mail address: shuyan@iastate.edu
canfield@ameslab.gov

CePd$_3$S$_4$ exhibits interplay between ferromagnetism (FM), quadrupolar order, and the Kondo effect at low temperatures with a FM transition temperature that is much higher than the value expected from the de Gennes scaling of the heavier RPd$_3$S$_4$ compounds. In this work, we investigated the electrical transport and magnetic properties of CePd$_3$S$_4$ under pressure up through 12 GPa so as to better understand the interplay between electronic and magnetic phases in this material. Our findings show that the low pressure FM state is suddenly replaced by a new magnetically ordered phase that is most likely antiferromagnetic that spans from ~ 7 GPa to ~ 11 GPa. Whereas this could be described as an example of avoided quantum criticality, given that clear changes in resistance and Hall data are detected near 6.3 GPa for all temperatures below 300 K, it is also possible that the change in ground state is a response to a pressure induced change in structure. The lack of any discernible change in the pressure dependence of the room temperature unit cell parameter/volume across this whole pressure range suggests that this change in structure is either more subtle than could be detected by our measurements (i.e. development of weak, new wave vector) or the transition is electronic (such as a Lifshitz transition).

## I. INTRODUCTION

Exotic states of matter and unusual physical properties are often found in the vicinity of quantum critical points (QCPs), or quantum phase transitions (QPTs). Whereas antiferromagnetic (AFM) transitions can often be suppressed to 0 K by a non-thermal control parameter such as pressure or magnetic field, thus reaching a QCP, current theoretical models suggest that, in the absence of disorder, ferromagnetic (FM) quantum criticality is generally avoided in favor of the transition becoming first order through a tricritical point, or the emergence of a new, spatially ordered phase such as AFM [1-9]. To date, experimental work on clean metallic FMs is largely consistent with this picture. Examples include MnSi [10-12], ZrZn$_2$ [13,14], CoS$_2$ [15,16], NiAl$_3$ [17], UGe$_2$ [18-20], and UTMX (TM= Co, Rh; X= Al, Ge) [21-24], among others, in which the FM transition changes from second- to first-order upon application of an external control parameter. Likewise, in other materials including LaCrGe$_3$ [25-27], La$_5$Co$_2$Ge$_3$ [28], Nb$_{1-y}$Fe$_{2+y}$ [29], and CeAgSb$_2$ [30], an AFM or spin density wave state emerges prior to quantum criticality. In the specific case of a non-centrosymmetric material with strong enough spin-orbit splitting energy of the conduction band near Fermi energy, recent theoretical work predicts a FM QCP may be accessible [31]. The recently discovered FM QCP in quasi-one-dimensional UIr [32,33], and CeRh$_6$Ge$_4$, [34,35] both of which have an easy axis anisotropy, could be an experimental example of this, albeit they were also understood as a quasi-one-dimensional FMQCP governed by the Kondo breakdown picture [36]. On the other hand, studies on heavily disordered FMs indicate that the transition can remain second order to the lowest temperatures accessible [37,38]. The diversity of possible scenarios in itinerant FMs, along with the unusual physical properties often found in the vicinity of the low temperature transitions, motivate the search for new metallic FMs with which to study the evolution of the magnetic state under the influence of applied pressure or magnetic field.

CePd$_3$S$_4$ is an underexplored and promising material. CePd$_3$S$_4$ crystallizes in the cubic NaPt$_3$O$_4$ structure with the space group *Pm*3*n* (space group #223) [39]. Here, the Ce$^{3+}$ has a quartet crystal field ground state ($\Gamma_8$ quartet) [40,41] that has both spin and orbital degrees of freedom and is well separated from the excited Kramer doublet ($\Gamma_7$) [42]. Consequentially, CePd$_3$S$_4$ exhibits both magnetic and orbital orders, undergoing a simultaneous FM and antiferroquadrupolar (AFQ) transition at $T_C$ = 6.3 K at ambient pressure [40,41,43].

Furthermore, the presence of hybridizing Ce$^{3+}$ ions make CePd$_3$S$_4$ a Kondo lattice compound, whose Kondo temperature, $T_K$ is comparable to ferromagnetic transition temperature, $T_C$ [43]. Interestingly, CePd$_3$S$_4$ shows a $T_C$ value of 6.3 K, which is significantly higher than the value of 0.07 K predicted from de Gennes scaling, and the $T_C$ of CePd$_3$S$_4$ is nearly the same as the $T_N$ = 5.8 K of GdPd$_3$S$_4$ [41]. Resonant X-ray results have suggested that whereas the quadrupole moments play a role of a primary order parameter in CePd$_3$S$_4$, the octupole moments play an important role as well. The coupling between the AFQ and antiferro-octupolar (AFO) orderings give rise to the complex magnetic structure of CePd$_3$S$_4$ [44] and may explain the anomalously high $T_C$ [41-44].

From the standpoint of FM quantum criticality, it is unclear how the FM order would evolve in the presence of AFQ order, and if there are possible interactions, that, for example, might allow for a FM QCP. Even if such a QCP is indeed avoided, external pressure, as a control parameter, can continuously drive the system into other intriguing phases. Here we report the systematic studies of electrical transport, magnetic and structural properties under high pressure on CePd$_3$S$_4$ single crystals and the assembly of a detailed Temperature-pressure ($T$ - $p$) phase diagram. Our results show that the low temperature, low pressure, FM state found in CePd$_3$S$_4$ discontinuously changes into non-FM state above ~ 6.3 GPa that most likely has an AFM component, creating a two-dome feature in the T-p phase diagram. Our temperature and pressure dependent resistance and Hall data indicate that there appears to be a phase transition appearing near the same 6.3 GPa pressure for temperature as high as 300 K all the way down to 10 K, suggesting that the two domes in the T-p phase diagram are less an avoided FM QCP and more a response of the system to a pressure induced change in electronic structure.

## II. METHODS

*Crystal growth and ambient pressure transport and magnetic properties*

We devised an optimized two-step process to grow large crystals of CePd$_3$S$_4$ [45] using fritted crucible sets. [46-47] First, a nominal composition of Ce$_5$Pd$_{58}$S$_{37}$ was loaded into an alumina crucible set and sealed in a fused silica tube. The tube was heated to 1150 °C, held for 8 hours, and cooled over 36 hours to 1050 °C, upon which the liquid was decanted. The tube was opened, and all solid material, mostly polycrystalline Ce$_2$S$_3$, was discarded and the captured decanted liquid reused in a new crucible set. The second crucible set was again sealed in silica and then warmed to 1075 °C. After holding for 8 hours, the furnace was slowly cooled over 150 hours to 900 °C, at which point the remaining solution was decanted. After cooling to room temperature, the tube was opened to reveal large, mirror-faceted crystals, a typical example of which is shown in the inset of FIG. 1(b). Powder X-ray diffraction (XRD) confirmed the products to be CePd$_3$S$_4$.

The magnetic properties of CePd$_3$S$_4$ at ambient pressure were measured using the Quantum Design Magnetic Properties Measurement System (MPMS-5) with magnetic field along [110] direction. Note, this particular, older, MPMS unit does not have a continuous low temperature (CLT) option. The consequences will be briefly discussed below, when magnetization measurements under pressure will be shown. The H/M(T) data are shown in FIG. 1(a). The curve above 100 K could be well fitted by a modified Curie Weiss law (MCW) [48],

$$\chi(T) = \frac{M(T)}{H} = \frac{C}{T - \Theta} + \chi_0 \qquad (1)$$

where $\chi_0$ is a temperature independent contribution to susceptibility, either core diamagnetism, Pauli paramagnetism, and/or van Vleck paramagnetism. C is Curie constant, and Θ is Curie Weiss temperature. MCW fitting gives a Curie Weiss temperature, Θ, around -21 K, and an effective moment $\mu_{eff}$ around 2.2 $\mu_B$/Ce. The measured $\mu_{eff}$ is close to the free Ce$^{3+}$ ion, 2.54 $\mu_B$/Ce, indicating Ce ions are in Ce$^{3+}$ state at high temperature. The negative Curie Weiss temperatures suggest a dominating AFM interaction, which is different from the suggested FM ground state [43], but meanwhile, supports the complex magnetic structure with multipolar ordering [44].

Despite the negative Θ value inferred from the high temperature MCW fits, CePd$_3$S$_4$ orders ferromagnetically at low temperatures. In FIG. 1(c), the $M(T)$ curve, under an applied field of 0.05 kOe along [110] direction, shows a steep increase below the Curie temperature, $T_C$, of ~ 6.5 K, indicating FM ordering. The criterion for defining $T_C$ used here is the intersection of two extended lines along the $M(T)$ curve above and below the transition temperature. A small applied magnetic field of approximately 5 kOe rapidly saturates the $T = 2$ K magnetization to a value just below 0.8 $\mu_B$/Ce, a small hysteresis loop (see FIG. 1(a) top left inset) is also observed. The magnetic properties are consistent with the findings presented in the previous report [43].

The electrical properties of CePd$_3$S$_4$ were measured by the Quantum Design Physical Properties Measurement System (PPMS) at ambient pressure. FIG. 1(b) presents the temperature dependence of resistance for single-crystalline CePd$_3$S$_4$ at ambient pressure. The behavior is metallic, and there is a Kondo minimum at, $T_{min}$ ~ 17 K, followed by an upturn upon further cooling until the local resistance maximum ($T_{max}$ ~ 7.5 K). The resistance upturn is likely due to Kondo screening by the conducting electrons, suggesting that CePd$_3$S$_4$ is a Kondo lattice with a Kondo temperature comparable to $T_C$ [43]. The resistance markedly decreases below $T_C$, indicating a loss of spin disorder scattering of conduction electrons due to FM ordering. The transition temperature, $T_C$, is around 6.3 K, defined as the midpoint of the d$R$/d$T$ shoulder below the transition temperature (as shown in FIG. 1(c)), which is consistent with previous studies [43]. The criteria for $T_C$ give very close $T_C$ values from resistance and magnetization data. We thus use the same criteria for defining transition temperatures in discussion of the high pressure data in next sections.

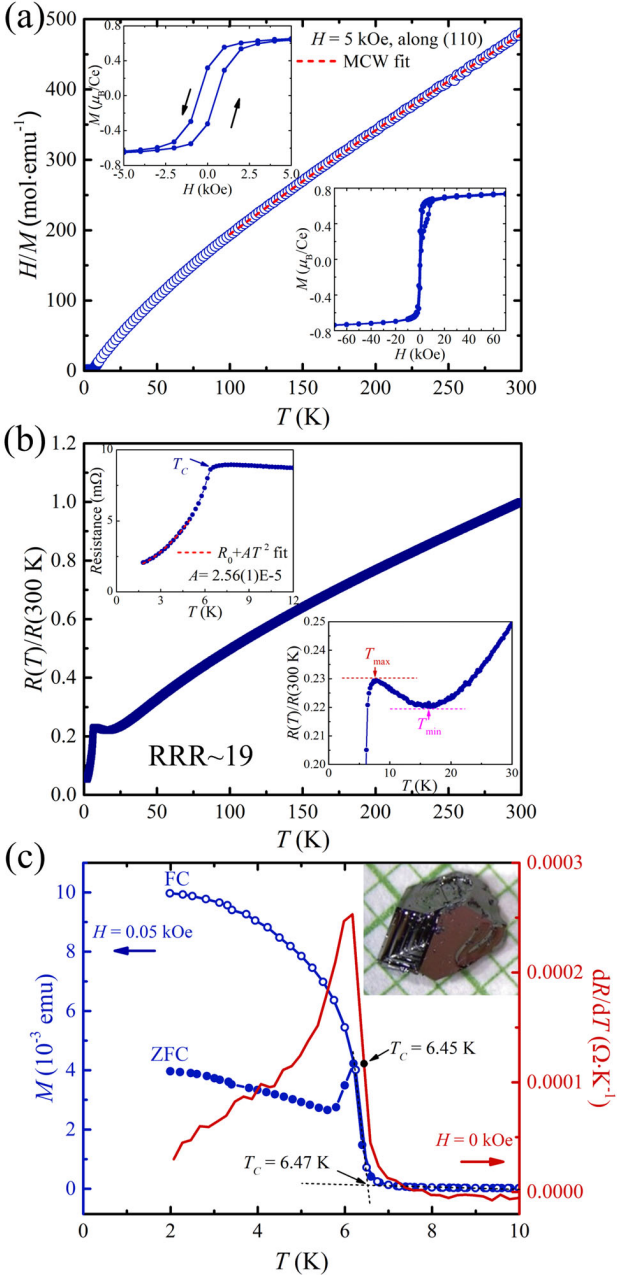

FIG. 1 Ambient pressure properties of $CePd_3S_4$. (a) Temperature dependence of $H/M(T)$ at 5 kOe. The top-left inset shows the low-field part of a four quadrant $M(H)$ loop. The bottom-right inset shows the $M(H)$ curve at 2 K with magnetic field up to 70 kOe. The red dashed curve represents the modified Curie Weiss fit (MCW) [48]. (b) Temperature dependence of normalized resistance ($R(T)/R(300K)$). The top-left inset shows the zoom-in plot of FM transition. The red dashed curve is fitted curve by Fermi liquid expression: $R(T) = R_0 + AT^2$. The bottom-right inset shows the zoomed-in plot of the local resistance minimum and maximum located at $T_{min}$ and $T_{max}$, respectively. (c) Zoom-in plot of the temperature dependence of: left hand axis: $M(T)$, and right hand axis: $dR(T)/dT$ to clearly show the FM transition. We used the intersection of two extended dashed lines along the $M(T)$ curve, and the mid-point of the $dR/dT$ shoulder as the criteria of the $T_C$. The inset shows the photograph of as-grown crystals on a mm grid.

### High pressure measurements

Electrical resistance was measured using the standard, linear four-probe configuration. Samples were measured in a piston-cylinder cell (PCC) [49] and a modified Bridgman anvil cell (MBAC) [50], using PPMS as a temperature and magnetic field platform. Pressure values for both cells, at low temperature, were inferred from the $T_C(p)$ of lead [51]. For the PCC, a 4:6 mixture of light mineral oil:n-pentane was used as the pressure medium, which solidifies, at room temperature, in the range of 3-4 GPa. For the MBAC, a 1:1 mixture of isopentane:n-pentane was used as the pressure medium, which solidifies, at room temperature, in the range of 6-7 GPa. Both solidification pressures are well above the maximum pressures achieved in the corresponding pressure cells, which suggests good hydrostatic conditions at 300 K during pressure changes [52,53].

To explore the phase diagram at higher pressure, resistance and DC magnetization measurements were performed down to temperatures as low as 1.8 K also using PPMS and MPMS, respectively, with commercial Diamond Anvil Cells [54,55]. Standard-design-type-Ia diamonds with a culet size of 700 μm were utilized as anvils. For electrical transport measurement the $CePd_3S_4$ single crystals were polished into thin flakes with a thickness of approximately 20 μm and cut into rectangular shapes about 100 μm x 50 μm for the electrical transport measurement. Platinum foil was used as electrodes to connect the sample, which was then loaded with a small ruby sphere into an apertured stainless-steel gasket covered with cubic-BN. To provide a more hydrostatic pressure environment, Nujol mineral oil was used as pressure transmitting medium, since: 1) fluid media, as opposed to solid media, can still maintain a quasi-hydrostatic pressure environment with a small pressure gradient below a liquid/glass transition [56-59]; 2) the use of a fluid medium avoids the direct contact between the sample and diamond culet which can lead to an additional uniaxial pressure component. For the resistance under applied magnetic field, we need to always consider the contributions of $R_{xx}$ and $R_{xy}$, so the magnetoresistance ($R_{xx}(T)$) is calculated by $(R(H, T) + R(-H, T))/2$, and Hall resistance ($R_H$) is calculated by $(R(H, T) - R(-H, T))/2$. The pressure was determined by ruby fluorescence [60,61] at room temperature.

For magnetization measurements, the sample was cut into dimensions of 200 μm x 200 μm x 40 μm and loaded with a ruby sphere into the hole of the apertured tungsten gasket. A high-purity non-magnetic tungsten sheet was used as the gasket. The background signal of the pressure cell (without the sample) at 0.2 kOe was measured under 3.4 GPa (which is closer to the pressure we observed an obvious change in magnetic ordering in the electrical resistance measurements (~6.3 GPa), so as to minimize the effect of the gasket-deformation and obtain a better resolution at high pressure). Then the DAC was opened and re-closed after loading the sample. The raw data of the pressure cell with and without the sample were analyzed based on the method outlined in ref [62]. Pressure was determined by ruby fluorescence [60,61] at room temperature.

To investigate the pressure effects on the crystal structure, the high pressure singe crystal x-ray diffraction (XRD) experiment was performed on a crystal of $CePd_3S_4$ with in-plane dimension of 0.037 × 0.031 mm$^2$ and less than 0.038 mm thickness up to 10.7 GPa. Prior to the high pressure experiment, the sample was mounted on a nylon loop with paratone oil and measured at ambient pressure to confirm its crystal structure. The sample was then loaded in the Diacell One20DAC [63] manufactured by Almax-easyLab with 500 μm culet-size extra aperture anvils. A 250 μm thick stainless steel gasket was pre-indented to 60 μm and a hole of 210 μm was drilled using an electronic discharge machining system. A 4:1 methanol–ethanol mixture was used as pressure transmitting medium. The pressure in the cell was monitored by the fluorescence R1-line of ruby [60,61]. The single crystal XRD measurements were performed using a Rigaku XtalLAB Synergy, Dualflex, Hypix single crystal X-ray diffractometer on Mo K$_α$ radiation ($\lambda$ = 0.71073 Å, micro-focus sealed X-ray tube, 50 kV, 1 mA), operating at room temperature. The total number of runs and images was based on the strategy calculation from the program CrysAlisPro 1.171.43.92a (Rigaku OD, 2023). Data reduction was performed with correction for Lorentz polarization. For the ambient pressure dataset, numerical absorption correction based on gaussian integration over a multifaceted crystal model Empirical absorption correction using spherical harmonics, implemented in SCALE3 ABSPACK scaling algorithm. The crystal structure was solved and refined using the Bruker SHELXTL [64,65] Software Package.

## III. RESULTS

*Electrical resistance measurement in PCC and MBAC*

To explore the relationship between pressure and magnetic ordering temperature, electrical resistance data on $CePd_3S_4$ single crystals were measured under high pressure using several different pressure cells. To carefully investigate the phase diagram at the lower-pressure range, $CePd_3S_4$ single crystals were examined using two different pressure cells: the piston-cylinder cell (PCC) and the modified Bridgman anvil cell (MBAC), ranging from 0 to 2.05 GPa and 1.81 to 4.96 GPa, respectively. The data from the PCC sample (shown in FIG. 2) displays consistent behavior up to 2.05 GPa, with no changes in the shape of the rapid resistance-drop associated with the FM transition. The onset temperature of the resistive feature that we associate with $T_C$ increases monotonically from 6.3 K at ambient pressure to 7.2 K at 2.05 GPa (as indicated in the inset of FIG. 2(a)). Similarly, in the MBAC measurements (shown in FIG. 3), the transition temperature $T_C$ increases in a monotonic fashion from 7.5 K at 1.81 GPa to 8 K at 3.68 GPa, which is qualitatively consistent with the PCC results (see FIG. 7(a) below for the full, composite $T$ - $p$ phase diagram assembled from all our diverse pressure cell measurements). However, the measured $T_C$ at 1.81 GPa in the MBAC (about 7.5 K) is higher than that in the PCC (approximately 7.1 K). This discrepancy may be attributed to an additional uniaxial pressure component [66].

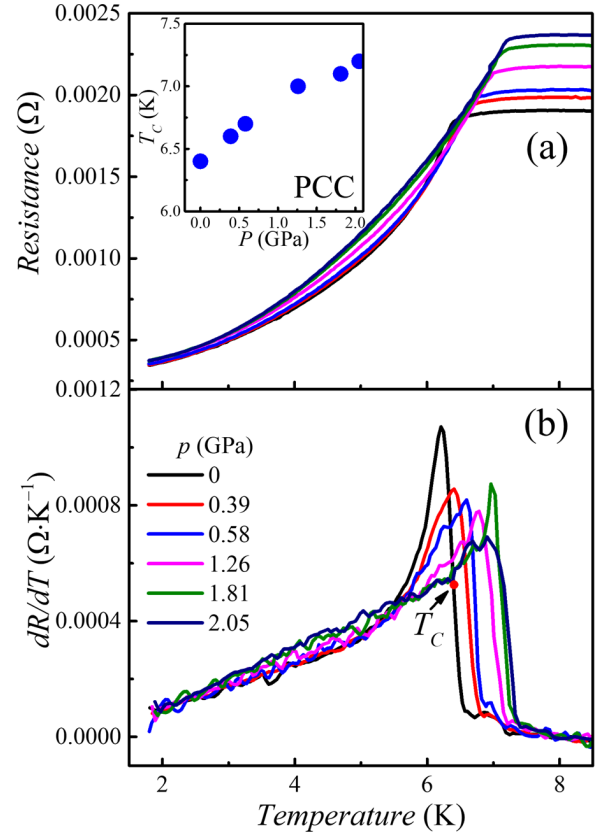

FIG. 2 High pressure resistance measurement in PCC. (a) temperature dependence of the resistance at various pressures. The inset shows the pressure dependence of the ferromagnetic $T_C$. (b) temperature dependence of d$R$/d$T$ at various pressures. Criterion for $T_C$ is the midpoint of the sharp rise in d$R$/d$T$. The red dot locates this midpoint for the ambient pressure data. (a) and (b) use the same legend. The resistance data over the full-temperature range can be found in the appendix FIG. A1(a).

As the pressure is further increased, $T_C$ decreases to approximately 7.5 K at 4.96 GPa (as indicated in the inset of FIG. 3(a)). The basic shape of the resistive feature stays the same up through the 4.35 GPa curve. For the 4.96 GPa data set the sample appears to have been damaged (see appendix, FIG. A1(b) for full data set), changing the measured resistance and, most likely, the connectivity and current path in the sample. We can still infer $T_C$ and its value is consistent with other $T$ - $p$ data, shown in FIG. 7(a) below.

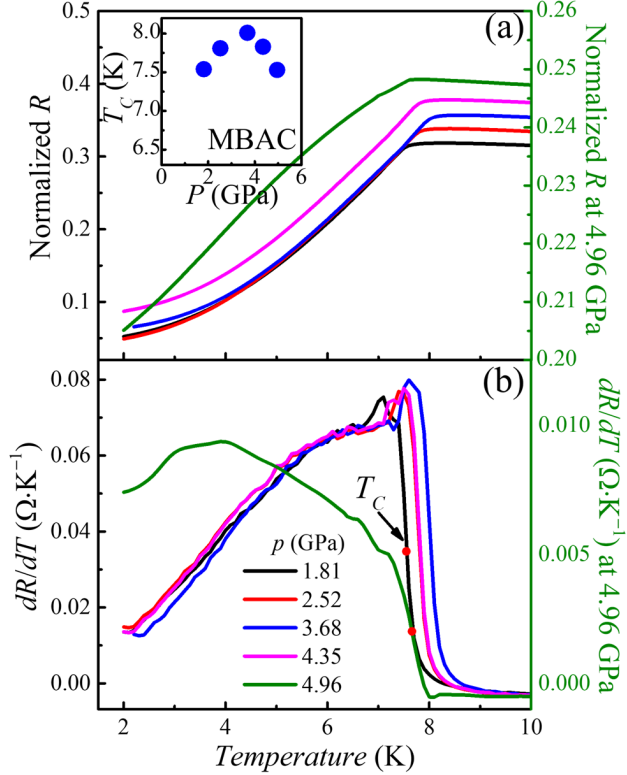

FIG. 3 High pressure resistance measurement in MBAC. (a) temperature dependence of normalized resistance ($R(T)/R(300K)$) at various pressures. The right hand axis corresponds to the data at 4.96 GPa. The inset shows the pressure dependence of ferromagnetic $T_C$. (b) temperature dependence of normalized $dR/dT$ at various pressures. The right hand panel shows the data at 4.96 GPa. (a) and (b) use the same legend. The resistance data in a full-temperature range can be found in FIG. A1(b). The same criterion for determination of $T_C$ as was used for the PCC data was used for the MBAC data and the inferred $T_C$s are shown by the red dots on the $p$ = 1.81 and 4.96 GPa data sets.

*Electrical resistance and DC magnetization measurements in DAC and MagDAC*

Upon analyzing the above results, we realized that studying the possible FM quantum criticality on CePd$_3$S$_4$ would require higher pressure to complete the phase diagram. To this end, we conducted electrical resistance and DC magnetization measurements up to 12.2 GPa and 7.7 GPa using DAC and MagDAC, respectively. FIG. 4 shows the resistance data collected using DAC. By comparing the resistance data at 1.81 GPa in PCC and MBAC, and at 1.6 GPa by DAC, shown in the appendix, FIG. A2, we find 3 different pressure cells show very similar results in both $R(T)$ and $dR/dT$. This suggests that data collected using different pressure cells can be safely compared, and that the $T$-$p$ phase diagram can be effectively constructed by combining all the data.

Examining the DAC data in FIG. 4 in greater detail, we see that as pressure is increased, we observed a dome-like $T_C$ behavior (FIG. 4(a), inset), where $T_C$ initially rises to approximately 8 K at 2.9 GPa before decreasing down to around 5.7 K at 6.3 GPa. These data are consistent with the PCC and MBAC results (see FIG. 7(a) below). Interestingly, by further increasing the pressure, we observed a second pressure dome from 6.3 GPa to 10.7 GPa (FIG. 4(b) and FIG. 4(d)). The transition can still be traced as a rapid resistance drop up to 9.0 GPa, which is similar to the resistance curve seen in the first pressure dome. For 10.7 GPa there is still a resistive feature, albeit much broader, for 12.2 GPa there is no discernible resistive feature that can be associated with a phase transition.

To investigate the ground states under the two pressure domes, the temperature dependence of resistance at various magnetic fields was studied. The results show that at low pressures, e.g., 1.6 GPa in FIG. 5(a), when the magnetic field is increased, the PM-FM transition broadens and shifts to higher temperatures. This is consistent with the resistive signature of FM transition in a metallic sample. However, at 8.22 GPa shown in FIG. 5(b), the transition temperature estimated from the resistive drop remains essentially constant with the external field up to 88 kOe. $R(T)$ curves under 0 field and 88 kOe for a wider range of pressures are shown in FIG. A3, which illustrate that $T_C$ at and below 6.3 GPa clearly increases with the magnetic field, whereas for 7.1 GPa – 10.7 GPa the resistive feature associated with the transition (maximum slope of loss of spin disorder scattering) doesn't change (or very slightly decreases) with the field up to 88 kOe. These observations provide a clear boundary between the FM and non-FM region at ~6.3 GPa. At and above 7.1 GPa, the behavior may indicate an AFM ground state that is robust to applied magnetic field, suggesting that the AFM order parameter may not be the primary order, or a combination of multipolar ordering with little or no staggered moment associated with it.

It is also noteworthy that above 7.1 GPa, the local resistance minimum in $R(T)$ curve that is thought to be associated with the Kondo screening at ambient pressure, becomes less pronounced, however, the temperature of the resistance minimum increases significantly, as shown in the appendix FIG. A1(c). The sudden rapid increase in $T_{min}$ may indicate the increase of the Kondo temperature due to the increasing of hybridization under high pressure.

To more directly investigate the ground state of the second pressure dome, high pressure measurements of temperature dependent magnetization were performed. FIG. 6 presents FC magnetization data taken in a 0.2 kOe magnetic field applied along [110] direction. The data show the same transition temperatures based on the criterion of the FM $T_C$ outlined above. As the pressure exceeds 6.1 GPa, the sharp FM transition changes to a broad feature with significantly reduced magnetization for $p$ = 6.4 GPa. For $p$ = 7.0 GPa this feature is barely discernible, and it is completely gone or not observable for 7.4 and 7.7 GPa.

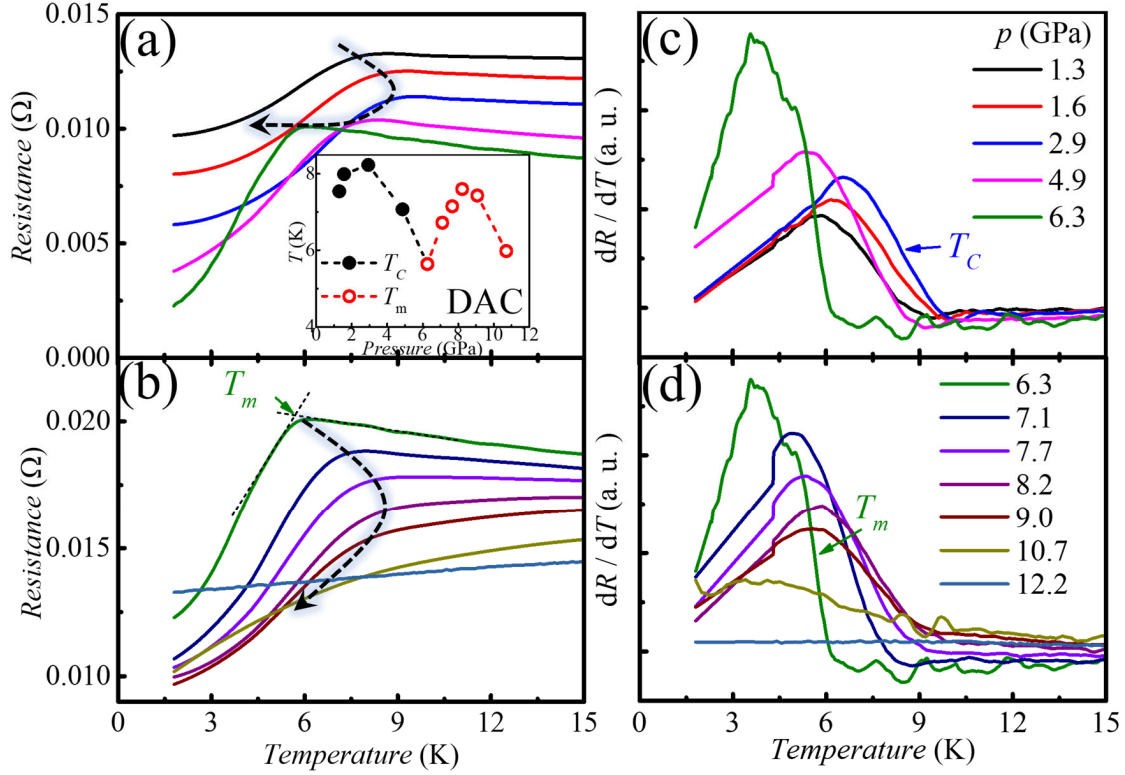

FIG. 4 High pressure resistance measurement in DAC. (a), (b) temperature dependence of resistance at various pressures. Each resistance curve is shifted to clearly see the evolution of the transitions. The inset shows the pressure dependence of resistance transition temperature. The black dashed curve with arrow is the guide to the eyes of how transition temperatures evolve with pressure. The dashed lines without arrows are the extended lines along the resistance curves showing the criterion of transition temperatures. (c), (d) temperature dependence of d$R$/d$T$ at various pressures. (a) and (c) share a common legend, as do (b) and (d). $T_C$, and $T_m$ are defined by the midpoints of the d$R$/d$T$ shoulders. All $R(T)$ curves are shifted so as to better show the evolution of $T_m$ with the pressure. The resistance data in a full-temperature range are shown in the appendix FIG. A1(c) below.

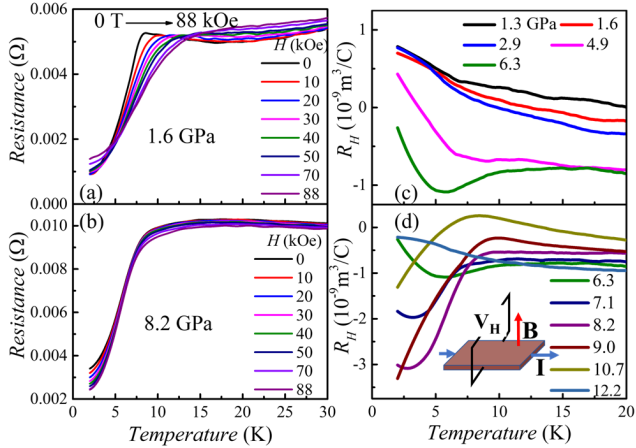

FIG. 5 (a), (b) temperature dependence of the resistance at various magnetic fields, at 1.6 and 8.22 GPa, respectively. (c), (d) temperature dependence of Hall coefficient $R_H$ at various pressure from 1.3 GPa to 6.3 GPa, and from 6.3 GPa to 12.2 GPa, respectively. A ± 88 kOe magnetic field was applied along [110] direction to measure the $R_H$. The inset in (d) shows the configuration of Hall measurement.

In addition, the temperature-dependent Hall coefficient ($R_H$) under pressure was studied to gain insights into the nature of the different transitions at high pressures, as shown in FIGs. 5(c), (d).

At lower pressures, the $R_H$ data, inferred from ± 88 kOe applied field data, exhibits a weak temperature dependence above ~10 K, and then a slight and broad transition-like upturn at lower temperature. This is consistent with the observations from FIG. 5(a), where an applied magnetic field broadens and shifts the ferromagnetic (FM) transition, resulting in a less distinct transition signature. The slight and broad upturn behavior might be attributed to the anomalous Hall effect resulting from the spontaneous magnetization of the FM order [67], based on the empirical formula,

$$R_H = R_0 H + R_S M \qquad (2)$$

where $H$ is magnetic field, $R_S$ is the anomalous Hall coefficient, and $M$ is the magnetization, which is shown in FIG. 6, albeit for much smaller fields.

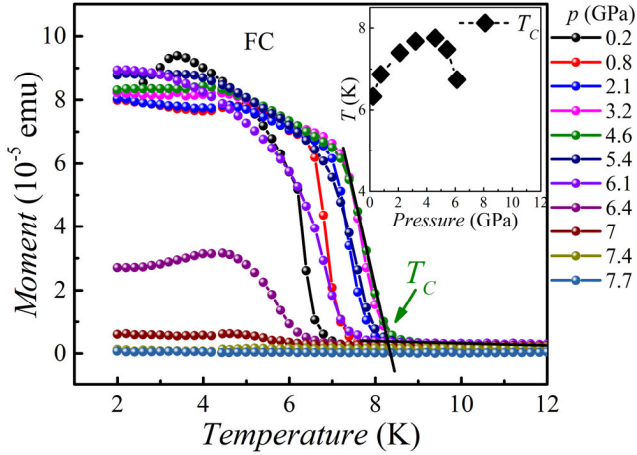

FIG. 6 High pressure magnetization measurement in MagDAC. The field cooling (FC) $M(T)$ measures at 0.2 kOe with field direction along [110] at various pressures. The black dashed lines demonstrate the criterion of defining $T_C$ (intersections of two extended lines along $M(T)$ data). Inset shows the pressure dependence of the transition temperature.

In the high-pressure dome region (7.1-10.7 GPa), a much clearer (and sharper) reduction in $R_H$ becomes apparent as the temperature decreases below the transition temperature (second dome phase line shown in FIG. 9(a)).

## IV. DISCUSSION

Based on the data shown in FIGs. 2-6, it becomes clear that there are two distinct, low temperature regions, a low pressure ($p \leq 6.3$ GPa) region in which we find FM order and a higher pressure region (6.3 GPa < p < 12 GPa) in which there is a new state that most likely has a AFM component. Notably, the $R(T)$ and $R_H(T)$ data presented in FIGs. 4 and 5 may suggest that distinct low pressure ($p < 6.3$ GPa) and high pressure ($p > 6.3$ GPa) regimes persist even above the $T < 9$ K transitions we have examined so far. In FIG. 7, the pressure dependence of $R$ and $R_H$ are shown for different temperatures. The resistance data show a very clear break in $R(p)$ isotherms at 6.3 GPa for temperatures up to room temperature. The Hall data show a very clear minimum centered at the same, 6.3 GPa, pressure for $T = 10$, 15, and 20 K. These data suggest that there is some clear change in the electronic states occurring near 6.3 GPa and extending up to 300 K.

Given that the feature in $R(p)$ is very clear at 300 K, we performed single crystal XRD at room temperature up through 10 GPa. We find that $CePd_3S_4$ maintains the same cubic structure ($Pm3n$, #223) over the whole pressure range (see FIGs. A4 and A5 in the appendix) with a roughly 9.4% decrease in the unit cell volume between 0 and 10 GPa (FIG. 8). The $V(p)$ curve can be well fit with a 2$^{nd}$ order Birch-Murnaghan Equation [69]:

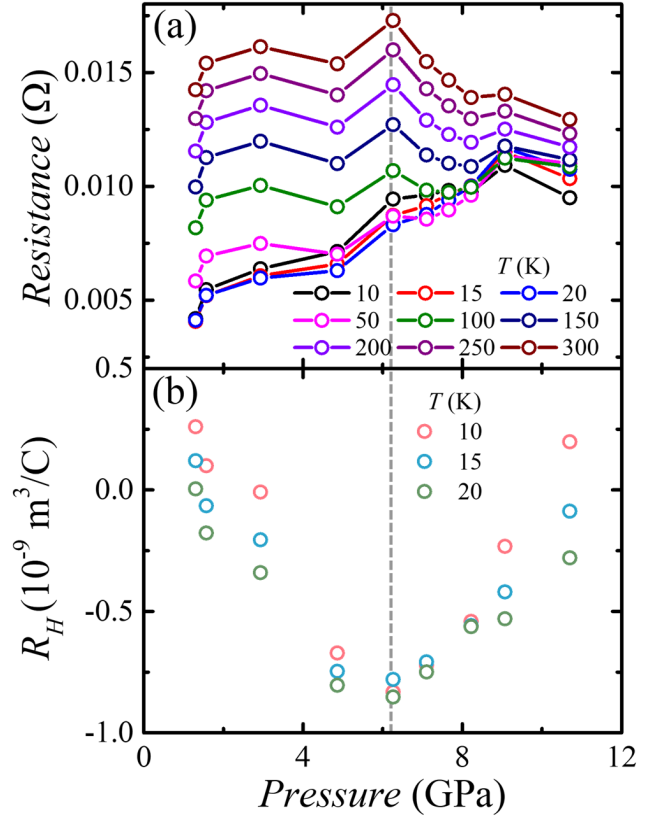

FIG. 7 Pressure-dependence of (a) Resistance ($R$) across a temperature range from 10 K to 300 K, and (b) Hall coefficient ($R_H$) at specific temperatures (10 K, 15 K, and 20 K). The black dashed line serves as a visual guide to the observed pressure (~6.3 GPa), highlighting local maxima in $R$ and minima in $R_H$.

$$p = -2\frac{9}{2}K_{T_0}V_0\frac{1}{2}\left[\left(\frac{V_0}{V}\right)^{\frac{2}{3}} - 1\right]\left[-\frac{1}{3V_0}\left(\frac{V_0}{V}\right)^{\frac{5}{3}}\right]$$
$$= \frac{3}{2}K_{T_0}\left[\left(\frac{V_0}{V}\right)^{\frac{7}{3}} - \left(\frac{V_0}{V}\right)^{\frac{5}{3}}\right] \quad (3)$$

with the isothermal bulk modulus $K_{T_0} = 109(2)$ GPa, which is similar to that of $EuPd_3S_4$ [68], and the volume at zero pressure $V_0 = 302.8(2)$ Å$^3$, further suggesting that $CePd_3S_4$ does not undergo any dramatic or conspicuous structural phase transition. The data presented in FIG. 8, as well as FIGs. A4 and A5 in the appendix, indicate that whereas there is not a sudden discontinuity in the lattice parameter as the pressure passes through 6.3 GPa, there may well be some subtler change in structure that still needs to be identified by more detailed, e.g. synchrotron, measurements.

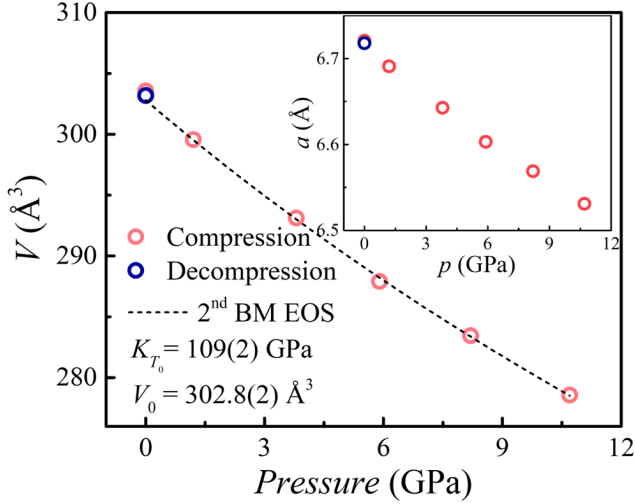

FIG.8 Pressure dependence of the unit cell volume ($V$). The black dashed curve shows the fitting by 2$^{nd}$-order Birch-Murnaghan equation of state (2$^{nd}$ BM EOS). The inset shows the pressure dependence of the lattice parameter ($a$).

Fig. 9(a) presents the $T$-$p$ phase diagram for CePd$_3$S$_4$. The left hand axis for $T \leq 10$ K presents the FM and AFM phase lines that we were able to determine from the data presented in FIGs. 2-4, and 6; the left hand axis for $T \geq 10$ K (log scale) indicates where we find the break in behavior in the $R_H(p)$ and $R(p)$ data. There is a very clear, essentially vertical line at ~6.3 GPa extending from 300 K all the way down to base temperature, dividing the $T$-$p$ phase diagram in the ordered state into low-pressure FM and high-pressure AFM phases, respectively. Given that we do not detect any conspicuous change in crystal structure or see any break in the pressure dependence of the unit cell volume, as we cross 6.3 GPa at 300 K, it is likely that there is an electronic transition, such as a Lifshitz transition or some subtle change in crystallographic structure that we cannot detect, taking place. The change in ground state ordering at this pressure, then, is less an example of "avoided FM quantum criticality" and rather the response of the system to change in the electronic states near the Fermi-level.

The $R(T)$ data below the magnetic ordering temperature from ambient pressure to 9.0 GPa are analyzed using the Fermi-liquid expression,

$$R(T) = R_0 + AT^2 \quad (4)$$

as depicted in FIGs. A6 and A7. The fits are good at temperature ranges from 1.8 K to up to approximately 4 K. On the other hand, a more general power law function,

$$R(T) = R_0 + AT^n \quad (5)$$

is employed to the low temperature $R(T)$ curves at and above 10.7 GPa due to the obvious deviations from the $AT^2$ fitting. The results show that n is very close to 1 at and above 10.7 GPa. As shown in FIG. 9(b) and FIGs. A6 and A7 in the appendix, there is a discontinuity in the prefactor (A) of $T^2$ resistivity near the 6.3 GPa pressure and what might be a region with non-Fermi liquid (NFL) like behavior on the high pressure side of the second dome, suggesting that there could be an actual quantum critical point in the 10 to 12 GPa pressure range. This possible NFL behavior would be associated to the single stage Kondo effect and suggesting a strong coupling between the local moment and the itinerant electrons. On one hand, this can be consistent with the Ce still being fundamentally trivalent in nature. Clearly at some, most likely much higher pressure, there should be a valence collapse into the tetravalent state. On the other hand, the NFL behavior could be due to the coupling between the quadrupolar moments and the conduction electrons. For example, in specific systems such as UBe$_{13}$ [70] and PrTi$_2$Al$_{20}$ [71], the transition into unconventional superconductivity from a non-Fermi liquid state is attributed to the multi-channel Kondo effect. Additional measurements, with greater data density between 10-12 GPa, at $p > 12$ GPa, and lower temperature below 1.8 K are needed to more clearly resolve the evolution of second dome and the entire physical picture of $CePd_3S_4$ system.

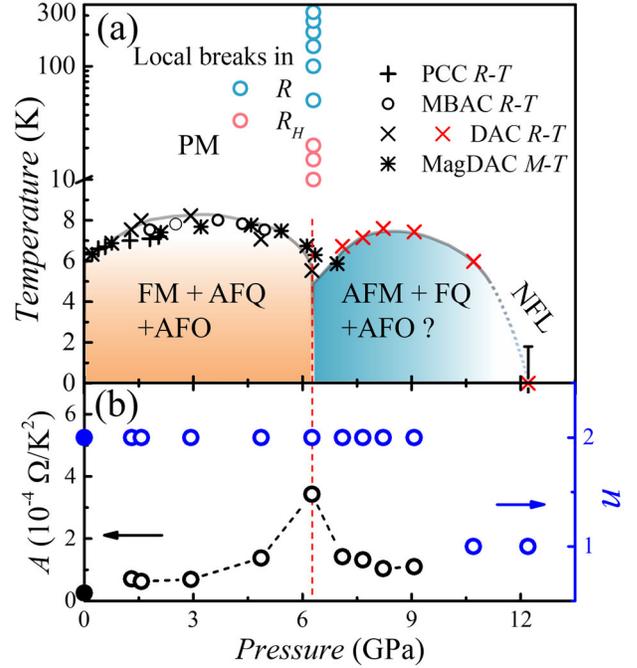

FIG. 9 Temperature-pressure ($T$-$p$) phase diagram (a) $T \leq 10$ K region depicts $T$-$p$ phase diagram of CePd$_3$S$_4$ as determined from resistance measurements by PCC, MBAC, $T \geq 10$ K region with y-axes in log scale records the pressure points at which the $R(p)$, or $R_H(p)$ show breaks (local maximum or minimum), and DAC, and the magnetization measurement by MagDAC. The vertical red dashed lines crossing all panels mark the lower and upper boundaries of two regions with different ground states at low temperatures: (0 - 6.3 GPa: FM + AFQ+ AFO) and (6.3 to ~12 GPa: antiferromagnetic (AFM)+ FM + AFO). (b) pressure dependence of fitting parameters A (left panel) and n (right panel) with the low temperature resistance data (from 1.8 K to ~ 4 K) fitted by power law, $R = R_0 + AT^2$, when $p < 10.7$ GPa, and $R = R_0 + AT^n$, when $p \geq 10.7$ GPa. For comparison, the black and blue solid circles are corresponding values fitted from the R(T) curve measured at ambient pressure outside the DAC (shown in FIG. 1(b)).

The near vertical nature of the ~ 6.3 GPa phase line merits further discussion. Given that in FIG. 8 we see that there is a 9.4% decrease in the room temperature unit cell volume between ambient pressure and 12 GPa, and given that for $EuPd_3S_4$ (in lieu of the data for $CePd_3S_4$) there is a reported ~ 0.4% decrease in the unit cell volume between room temperature and 10 K [72], there is a change in volume upon cooling that is equivalent to a half GPa change in pressure. Based on the assumption that the ~ 6.3 GPa transition takes place at a given volume of the cubic unit cell, this suggests that much tighter pressure spacings in the 6 - 7 GPa range *might* allow for the detection of this line via temperature dependent measurements. Conversely, given that the resistance and Hall data shown in FIG. 7 have pressure spacings of roughly 1 GPa, it is not surprising that such a narrow pressure window was missed. As a final comment on this topic, even if the hypothetical critical volume could be passed by cooling through some temperature, say 100 K, then, given the steepness of this phase line, any features that might exist in the $R(T)$ or $R_H(T)$ data may well be very broad.

Although the P ~ 6.3 GPa change in low temperature state appears to be associated with pressure induced electronic transition or some subtle change in crystallographic structure that we didn't detect, we can still try to understand the relationship between the ordered states. We know from the ambient pressure studies [43,44] that the lower pressure dome is AFQ + AFO + FM. The higher pressure dome manifests as a transition-like feature in the R(T) data and exhibits minimal change under magnetic fields up to 88 kOe. The Hall resistivity has a minimum at the 6.3 GPa transition between the two, which suggests a change in the band structure, but does not show the step-like behavior expected for a Kondo breakdown transition. Therefore, it is likely that the new ordered phase is also a type of $\Gamma_8$ multipolar order. In terms of $J_z$ eigenstates ($J_z$ as angular moment operator where z-axis is parallel to one of the fourfold axes of cubic symmetry), the $\Gamma_8$ quartet is,

$$|\Gamma_8\ a, \pm\rangle = \sqrt{\frac{5}{6}}\left|\pm\frac{5}{2}\right\rangle + \sqrt{\frac{1}{6}}\left|\mp\frac{3}{2}\right\rangle\ |\Gamma_8\ b, \pm\rangle = \left|\pm\frac{1}{2}\right\rangle \quad (5)$$

Where, the *a* and *b* "doublets" have opposite $O_2^0$ quadrupolar moments, and while both have dipolar moments, the *a* doublet has a larger dipole moment. If we assume there is two Ce atoms per unit cell in the ordered phase, the low pressure AFQ/AFO/FM phase has $|\Gamma_8\ a, +\rangle$ and $|\Gamma_8\ b, +\rangle$ states, which correspond to staggered $O_2^0$ quadrupolar moments and $T_z^\alpha$ octupolar moments. ※

The high pressure phase clearly has no net moment and so must be AFM. If we assume that there are still two sites per unit cell, we must have ferroquadrupolar (FQ) order and accompanying AFO order. The most likely candidate, given the lack of field response, is $|b +\rangle$, $|b -\rangle$ on alternating sites, which has staggered small ($J_z = \pm 1/2$) moments, FQ order (and an associated lattice distortion) and AFO without any net octupolar moments.

## V. CONCLUSION

In summary, our electrical transport and DC magnetization measurements of $CePd_3S_4$ under high pressure led to the discovery of a new non-FM phase above ~6.3 GPa (most likely AFM). The formation of this phase is most likely associated with a near vertical line in the *T* - *p* phase diagram that we associate with a pressure induced electronic phase transition or some subtle change in crystallographic structure that we are unable to detect. On further pressure increase this new phase is suppressed at ~ 11 GPa and non-Fermi liquid behavior is observed above 11 GPa. Further studies of $CePd_3S_4$ are needed to more clearly describe the nature of the ~ 6.3 GPa transition as well as to examine the upper quantum phase transition for further evidence of possible quantum critical effects and possible, further emergent phases.

## VI. ACKNOWLEDGEMENT

Work at Ames National Laboratory is supported by the US DOE, Basic Sciences, Material Science and Engineering Division under contract no. DE-AC02-07CH11358. T.J.S. was supported in part, by the Center for Advancement of Topological Semimetals (CATS), an Energy Frontier Research Center funded by the U.S. Department of Energy Office of Science, Office of Basic Energy Sciences, through Ames National Laboratory under its Contract No. DE-AC02-07CH11358 with Iowa State University. W. X. and H-Z W are supported by the U. S. Department of Energy (DOE), Office of Science, Basic Energy Sciences under award DE-SC0023648.

---

※ *There is a technical complication that both the magnetic dipole, $\vec{J}$, and relevant octupole, $\vec{T}^\alpha$, are $\Gamma_{4u}$ irreps and mix with one another to form $\vec{\sigma}$ and $\vec{\eta}$ order parameters that are uniform and staggered, respectively, but the dipolar and octupolar moment sizes vary from site to site, with their signs being uniform and staggered, respectively [73,74].*

# Appendix

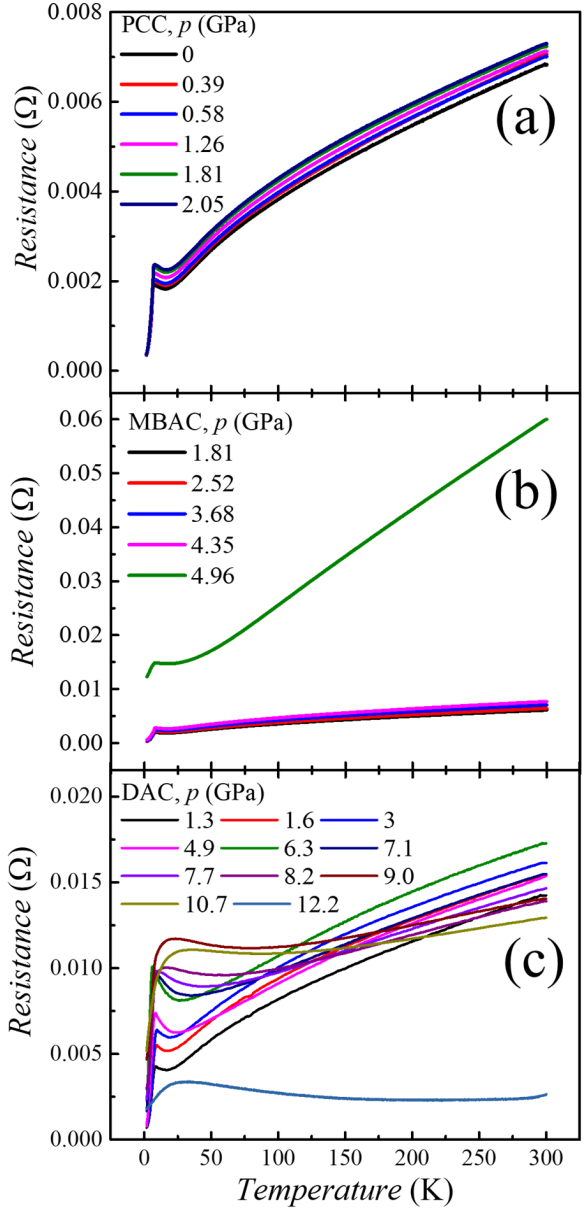

FIG. A1 Resistance as a function of temperature at various pressures measured in (a) PCC, (b) MBAC, and (c) DAC.

FIG. A1 shows full temperature-range R(T) curves measured in 3 different pressure cells as supporting information of FIGs. 1-3.

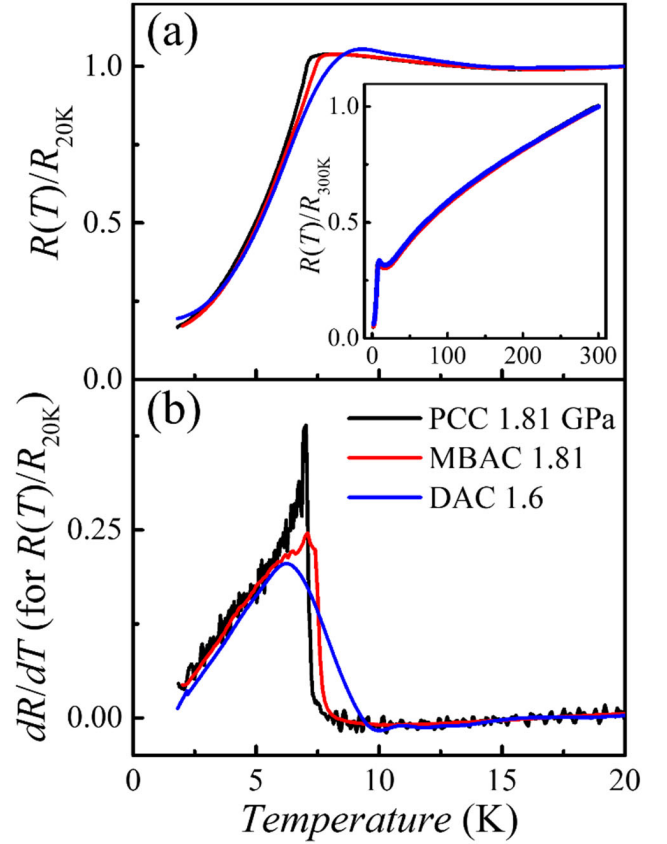

FIG. A2 Comparison of the data collected from PCC at 1.81 GPa, MBAC at 1.81 GPa, and DAC at 1.6 GPa on (a) $R(T)/R(20\,K)$ and (b) $dR(T)/dT$ curves on $R(T)/R(20\,K)$ data. Inset of (a) is the $R(T)/R(300\,K)$ curve in a full temperature range. (a) and (b) share the same legend.

FIG. A2 shows that R(T) curves measured in 3 different pressure cells, at similar pressures, give very similar results.

In addition to the $R(T)$ curves at various magnetic fields at 1.6 GPa and 8.2 GPa shown in FIGs. 5 (a) and (b) respectively, we also show $R(T)$ at zero applied field and 88 kOe magnetic field at all pressures in FIG. A3. We can clearly see the resistance transition broadens and shifts to higher temperature in the lower pressure region, or dome, (1.3 GPa to 6.3 GPa), and has only minor changes with field in the upper pressure region, or dome, (8.2 to 10.7 GPa).

FIGs. A4 and A5 present the x-ray data taken under pressure at room temperature that were solved to provide the data presented in FIG. 8 in the main text.

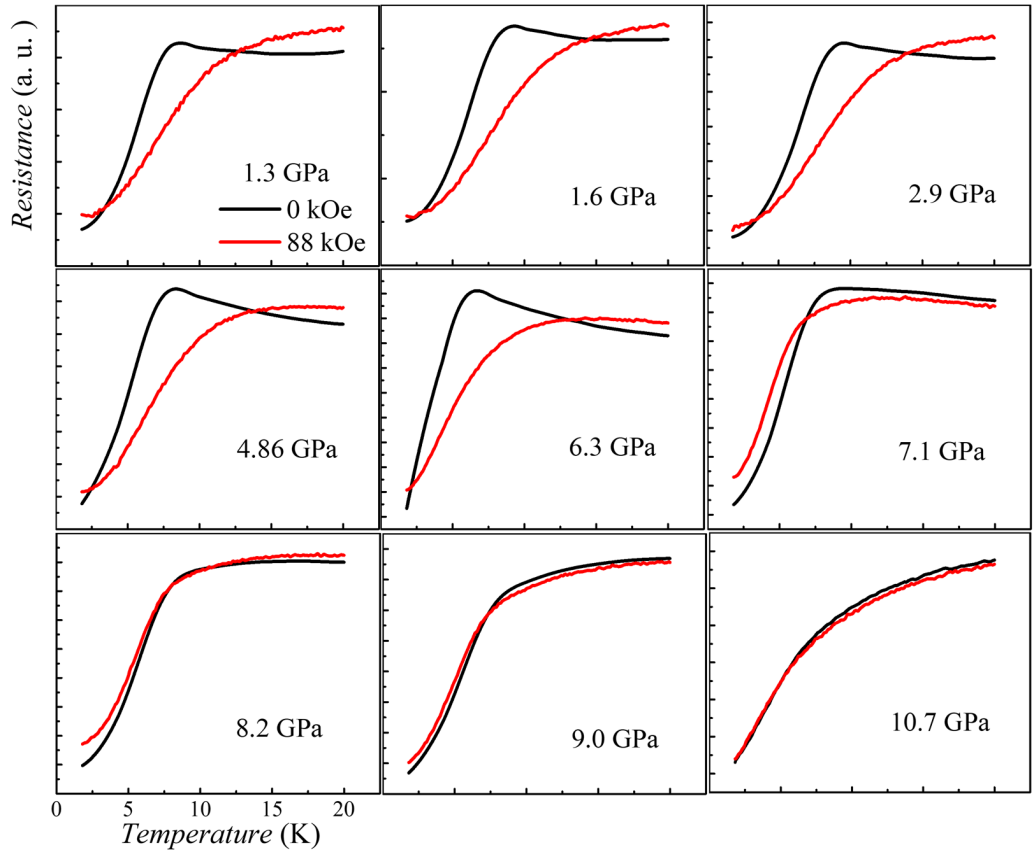

FIG. A3 Resistance as a function of temperature at zero field and 88 kOe under various pressures.

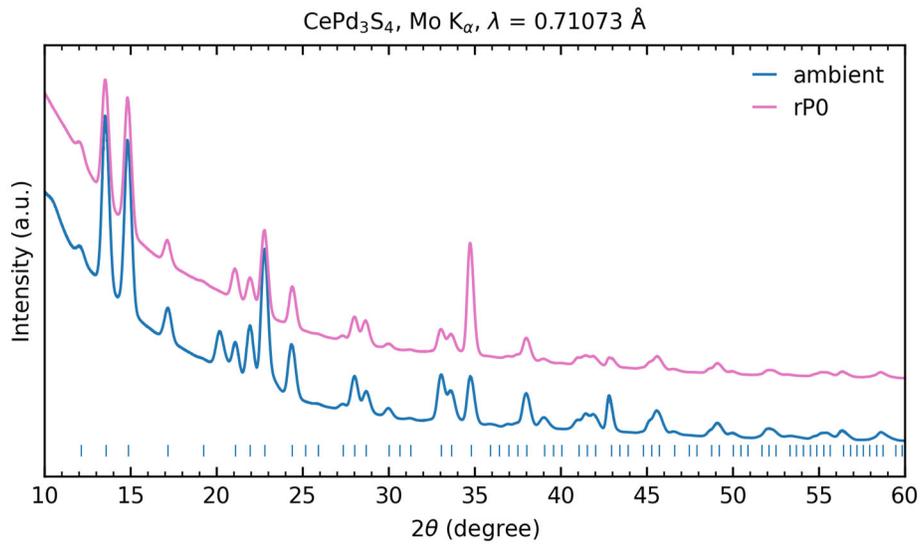

FIG. A4 The powder XRD pattern generated from 2D images obtained in the single crystal XRD measurements at the initial ambient pressure (blue) and ambient pressure after decompression (pink). Extra diffraction peaks come from the iron and diamond, which are clearly shown in FIG. A5

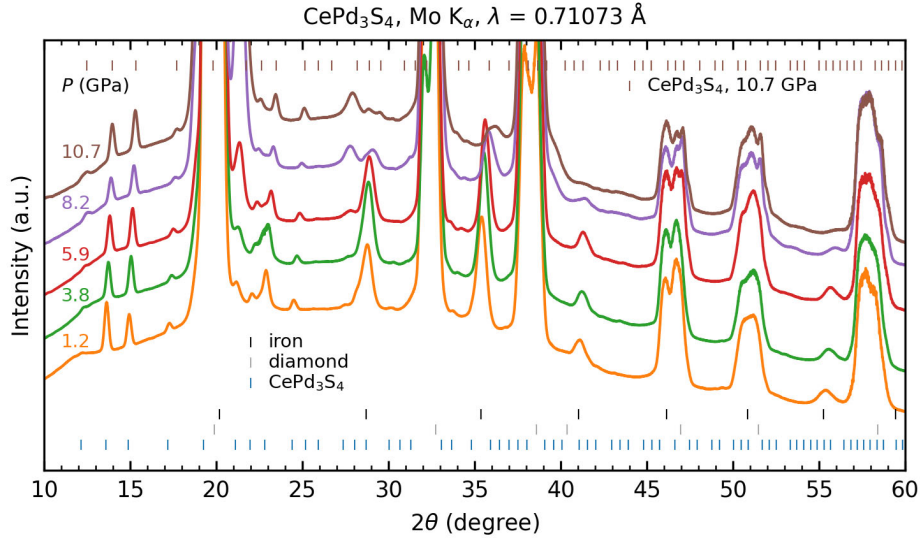

FIG. A5 Pressure dependent powder XRD pattern generated from 2D images obtained in the single crystal XRD measurements.

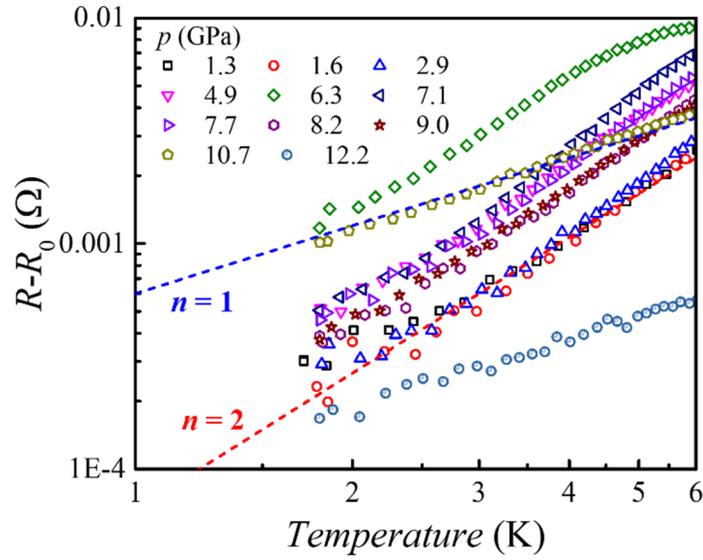

FIG. A6 $R-R_0$ as a function of temperature in log-log scale at various pressures. The blue and red dashed lines are guides showing slopes for various low temperature exponents, n = 1, and 2, respectively. Here, the $R_0$ was determined by fitting a power-law function to the R(T) curve, which is shown in FIG. A7.

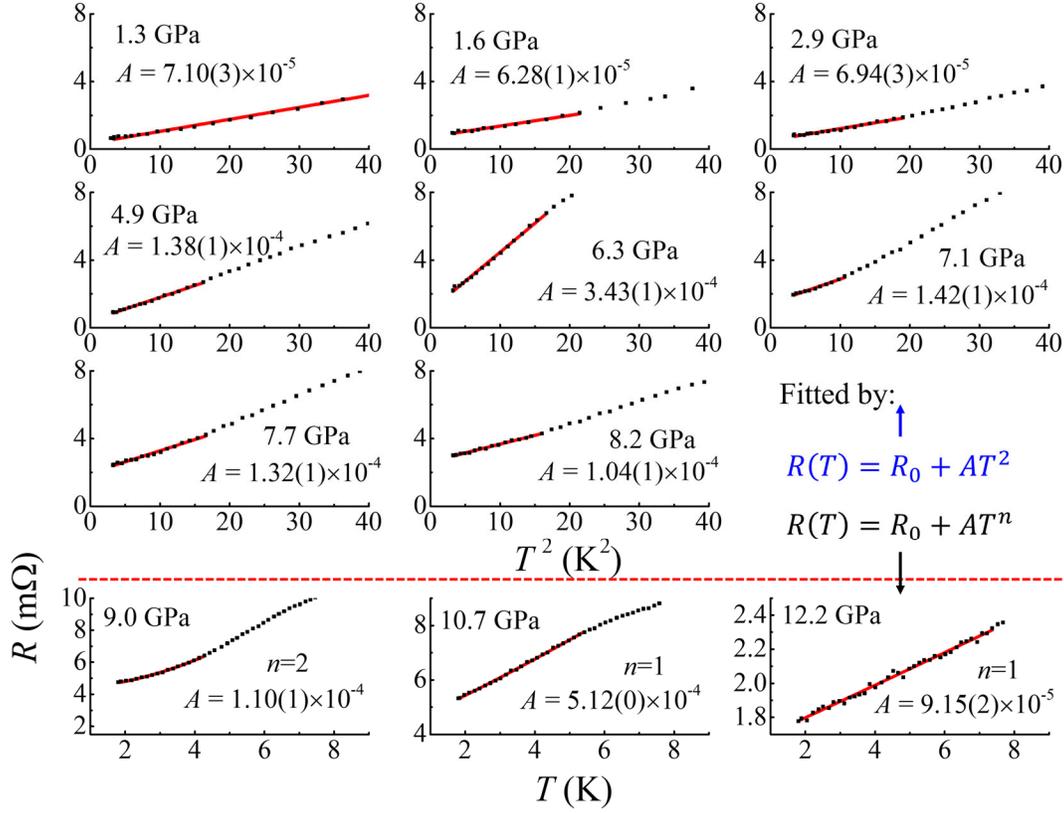

FIG. A7 Power law fitting results on $R(T)$ curves at various pressures. All results from 1.3 GPa to 8.2 GPa are plotted in the same $R$-$T^2$ scales. Results from 9.0 GPa to 12.2 GPa are plotted in $R$-$T$ scales. Black dots are the measured data and red solid lines are fitted curves. The 9 curves from 1.3 GPa to 8.2 GPa are fitted by Fermi liquid expression: $R(T) = R_0 + AT^2$, ($n=2$). The bottom 3 curves from 9.0 GPa to 12.2 GPa are fitted by more general power law: $R(T) = R_0 + AT^n$ and get $n$ factor as 2, 1 and 1.

In FIGs. A6 and A7, we show that the low temperature $R(T)$ curves below 9.0 GPa could be well fitted by power low with n factor of 2, whereas, at and above 10.7 GPa, the low temperature resistance behavior linearly down to 1.8 K, indicating a emergence of non-Fermi liquid behavior at high pressure.